\documentclass[aps,pre,twocolumn,amsfonts,amsmath,showpacs,a4paper]{revtex4}
\usepackage{color}
\usepackage{enumerate}
\usepackage{graphicx}
\usepackage{times}
\usepackage{float}

\newcommand{\eg}[1]{{\it e.g.\/}\ifx#1.\else\expandafter#1\fi}
\newcommand{\ie}[1]{{\it i.e.\/}\ifx#1.\else\expandafter#1\fi}

\newcommand{\df}[2]{\frac{\partial #1}{\partial #2}}

\newcommand{\e}{e}                 
\newcommand{\eqlabel}[1]{\label{eq:#1}}
\newcommand{\eq}[1]{\eqref{eq:#1}}
\def\eqreftwo(#1,#2){(\ref{eq:#1},\ref{eq:#2})}
\newcommand{\eqtwo}[1]{\eqreftwo(#1)}

\DeclareMathOperator{\Heav}{\Theta}     
\renewcommand{\i}{i}               
\newcommand{\Mx}[1]{\left[\begin{array}{cc}#1\end{array}\right]}
\newcommand{\mx}[1]{\mathbf{#1}}
\renewcommand{\Re}[1]{\mathop{\mathrm{Re}}\left(#1\right)}
\DeclareMathOperator{\Tr}{Tr}

\newcommand{\A}{\mx{A}}            
\renewcommand{\a}{a}               
\newcommand{\ac}{a_*}              
\newcommand{\ale}{A}               
\newcommand{\ble}{B}               
\newcommand{\C}{C}                 
\newcommand{\cg}{c}                
\newcommand{\Du}{D_u}              
\newcommand{\Dv}{D_v}              
\newcommand{\decr}{\mu}            
\newcommand{\dt}{\Delta t}         
\newcommand{\dx}{\Delta x}         
\newcommand{\eps}{\varepsilon}     
\newcommand{\evec}{\mx{v}}         
\newcommand{\evals}{\nu}           
\newcommand{\evalt}{\lambda}       
\newcommand{\f}{f}                 
\newcommand{\freq}{\omega}         
\newcommand{\g}{g}                 
\newcommand{\hone}{h_1}            
\newcommand{\htwo}{h_2}            
\newcommand{\ki}{k_1}              
\renewcommand{\L}{L}               
\newcommand{\n}{n}                 
\newcommand{\Period}{T}            
\renewcommand{\t}{t}               
\renewcommand{\u}{u}               
\newcommand{\umin}{\u_{\min}}      
\newcommand{\umax}{\u_{\max}}      
\newcommand{\ut}{\u_*}             
\newcommand{\ur}{\u_0}             
\newcommand{\us}{u_s}              
\renewcommand{\v}{v}               
\renewcommand{\vr}{\v_0}           
\newcommand{\w}{w}                 
\newcommand{\waven}{k}             
\newcommand{\icw}{\delta}          
\newcommand{\x}{x}                 
\newcommand{\xs}{\x_s}             

\newcommand{\Fig}[1]{Fig.~\ref{fig:#1}}
\newcommand{\fig}[1]{fig.~\ref{fig:#1}}
\newcommand{\figs}[1]{figures~\ref{fig:#1}}
\newcommand{\figref}[1]{\ref{fig:#1}}

\newcommand{\seclabel}[1]{\label{sec:#1}}

\newcommand{\dblfigure}[3]{
  \begin{figure*}[htbp!]
  \includegraphics{#1}
  \caption[]{#2}
  \label{fig:#3}
  \end{figure*}
}
\newcommand{\sglfigure}[3]{
  \begin{figure}[htbp!]
  \includegraphics{#1}
  \caption[]{#2}
  \label{fig:#3}
  \end{figure}
}

\begin{document}

\title{Classification of wave regimes in excitable systems with linear
  cross-diffusion}

\author{M. A. Tsyganov}
\affiliation{
  Institute of Theoretical and Experimental Biophysics, 
  Pushchino, Moscow Region, 142290, Russia}
\author{V. N. Biktashev}
\affiliation{College of Engineering, Mathematics and Physical Sciences,
  University of Exeter, Exeter EX4 4QF, UK}
\date{\today}

\begin{abstract}
  We consider principal properties of various wave regimes in two
  selected excitable systems with linear cross-diffusion in one
  spatial dimension observed at different parameter values. This
  includes fixed-shape propagating waves, envelope waves,
  multi-envelope waves, and intermediate regimes appearing as waves
  propagating fixed-shape most of the time but undergoing
  restructuring from time to time. Depending on parameters, most of
  these regimes can be with and without the ``quasi-soliton'' property
  of reflection of boundaries and penetration through each other. We
  also present some examples of behaviour of envelope quasi-solitons
  in two spatial dimensions.
\end{abstract}


\maketitle

\section{Introduction}
\seclabel{introduction}

The progress in the study of self-organization phenomena in physical,
chemical and biological systems is dependent on study of generation,
propagation and interaction of nonlinear waves in spatially
distributed active, e.g. excitable, systems with
diffusion~\cite{ref1}. An important general property of such
systems is their ability to generate and conduct self-supported
strongly nonlinear waves of the change of state of the medium. The
shape and speed of such waves in the established regime does not
depend on initial and boundary conditions and is fully determined by
the medium parameters. Until recently the results concerning such
systems have been focused on systems ``reaction+diffusion'' with a
diagonal diffusivity matrix, e.g. for two reacting components, 
\begin{equation}
  \df{\u}{\t} = \f(\u,\v) + \Du\nabla^2{\u}, \quad
  \df{\v}{\t} = \g(\u,\v) + \Dv\nabla^2{\v},                \eqlabel{RD}
\end{equation}
with non-negative diffusivities $\Du\ge0$, $\Dv\ge0$, $\Du+\Dv>0$.
However, a number of applications motivate consideration
  of a more generic class of reaction-diffusion systems, with
  non-diagonal elements of the diffusivity matrix
  (``cross-diffusion''), which can produce a number of unusual
  patterns and wave regimes, see e.g. a
  review~\cite{Vanag-Epstein-2009}. In this paper we concentrate on
  one subclass of such unusual wave regimes, which is associated with
  soliton-like interaction, i.e. penetration of waves upon impact with
  each other or reflection from non-flux boundaries. This is rather
  uncharacteristic of the waves in \eq{RD} with the exception of
  narrow parametric regions on the margins of the
  excitability~\cite{someref}. However, in systems with
  cross-diffusion, such ``quasi-soliton' behaviour can be observed in
  large parametric regions~\cite{QS1,QS6}. These phenomena have been
  observed in numerical simulations of two-component excitable media
  with cross-diffusion, both in linear formulation, e.g.
\begin{align}
  &
  \df{\u}{\t} = \f(\u,\v) + \Du\nabla^2{\u} + \hone\nabla^2\v, 
  \nonumber \\ &
  \df{\v}{\t} = \g(\u,\v) + \Dv\nabla^2{\v} - \htwo\nabla^2\u,           \eqlabel{RXD}
\end{align}
and nonlinear, ``taxis'' formulation,
\begin{align}
  &
  \df{\u}{\t} = \f(\u,\v) + \Du\nabla^2{\u}
    + \hone\nabla\left(u\nabla{\v}\right), 
  \nonumber \\ &
  \df{\v}{\t} = \g(\u,\v) + \Dv\nabla^2{\v}
    - \htwo\nabla\left(v\nabla{\u}\right),                  \eqlabel{RT}
\end{align}
where $\hone\ge0$, $\htwo\ge0$, $\hone+\htwo>0$. 

Quasi-solitons have similarities and differences with the classical
solitons in conservative (fully integrable) systems. The already
mentioned similarity is their ability to penetrate through each other
and reflect from boundaries. The differences are:
\begin{itemize}
\item The amplitude and speed of a true soliton depend on initial
  conditions. For the quasi-soliton, the established amplitude and
  speed depend on the medium parameters. 
\item The amplitudes of the true solitons do not change after the
  impact. The dynamics of quasi-solitons on impact is often naturally
  seen as a temporary diminution of the amplitude with subsequent
  gradual recovery.
\end{itemize}

Recently we have demonstrated ``envelope quasi-solitons'' in
one-dimensional systems with linear
cross-diffusion~\eq{RXD}~\cite{EQS}, which share some
phenomenology with envelope solitons in the nonlinear Schr\"odinger
equation (NLS) for a complex field $\w$~\cite{NLS},
\begin{equation}
  \i\df{\w}{\t} + \nabla^2\w + \w|\w|^2 = 0 .               \eqlabel{NLS}
\end{equation}
Namely, they have the form of spatiotemporal oscillations
(``wavelets’'') with a smooth envelope, and the velocity of the
individual wavelets (the phase velocity) is different from the
velocity of the envelope (the group velocity). This may be serious
evidence for some deep relationship between these phenomena from
dissipative and conservative realms. The link in this relationship is
cross-diffusion, which for NLS is revealed if is rewriten as a system
for two real fields $\u$ and $\v$ via $\w = \u - \i\v$ of the
form~\eq{RXD} with
\begin{align*}
  &
  \hone=\htwo=1, 
  \quad
  \Du=\Dv=0, 
  \\ &
  \f=\u(\u^2+\v^2), 
  \quad
  \g=-v(\u^2+\v^2). 
\end{align*}
Note the signs of the cross-diffusion terms in the componentwise form
of NLS and in \eq{RXD}.

Further investigation has revealed a great variety of the types
of nonlinear waves in excitable cross-diffusion systems. In this paper
we present some classification of the phenomenologies of such waves.

Our observations are made in two selected two-component kinetic
models, supplemented with cross-diffusion, rather than self-diffusion
terms; such terms may appear, say, in
mechanical~\cite{Cartwright-etal-1997},
chemical~\cite{Chung-Peacock-2007,Vanag-Epstein-2009}, biological and
ecological~\cite{UFN07,Murray-2003} contexts.  %
  We note that the case of \emph{only} cross-diffusion terms, with
  $\Du=\Dv=0$, is special in that the spatial coupling is then not
  dissipative, and all the dissipation in the system is due to the
  kinetic terms. So, theoretically speaking, this case may present
  features that are not characteristic for more realistic models.  In
  practice, however, these worries seem unfounded.  Parametric studies
  done in the past~\cite{QS1,QS2} indicate that the role of the
  self-diffusion coefficients $\Du$, $\Dv$ is not essential if they
  are small enough. Moreover, we have verified that the results
  presented below are robust in that respect, too.  In other words,
  regimes observed for $\Du=\Dv=0$ typically are qualitatively
  preserved, even if quantitatively modified, upon adding small $\Du$,
  $\Dv$. So in this study we limit consideration to $\Du=\Dv=0$ to
  reduce number of parameters and focus attention on effects of the
  cross-diffusion terms. %
Except where stated otherwise, the values of the
  cross-diffusion coefficients are $\hone=\htwo=1$.  We
consider the FitzHugh-Nagumo (FHN) kinetics,
\begin{equation}
  \f = \u(\u-\a)(1-\u) - \ki\v, \quad
  \g = \eps \u,                                             \eqlabel{FHN}
\end{equation}
for varied values of parameters $\a$, $\ki$, and $\eps$. 
As a specific example of a real-life system, we also consider the
Lengyel-Epstein (LE)~\cite{Lengyel-Epstein-1991} model of a
chlorite-iodide-malonic acid-starch autocatalitic reaction system 
\begin{equation}
  \f = \ale-\u - \frac{4\u\v}{1+\u^2}, \quad
  \g = \ble \left(\u - \frac{\u\v}{1+\u^2}\right) .         \eqlabel{LE}
\end{equation}
for varied values of parameters $\ale$ and $\ble$. 

\section{Methods}
\seclabel{methods}

We simulate \eq{RXD} in one spatial dimension for $\x\in[0,\L]$,
$\L\le\infty$, with Neumann boundary conditions for both $u$ and $v$.
We use first order time stepping, fully explicit in the reaction terms
and fully implicit in the cross-diffusion terms, with a second-order
central difference approximation for the spatial derivatives.  Unless
stated otherwise, we used steps $\dx=1/10$ and $\dt=1/5000$ for FHN
kinetics~\eq{FHN} and $\dx=0.1$ and $\dt=1/1000$ for LE
kinetics~\eq{LE}.

To simulate propagation ``on an infinite line'', we did the
simulations on a finite but sufficiently large $\L$ (specified in each
case), and instantanously translated the solution by $\delta\x_1=30$
away from the boundary each time the pulse, as measured at the level
$\u=\ut$, where $\ut=0.1$ for FHN kinetics and $\ut=1.5$ for LE
kinetics, approached the boundary to a distance smaller than
$\delta\x_2=100$, and filled in the new interval of $x$ values by
extending the $u$ and $v$ variables at levels $u=\ur$, $v=\vr$, where
$(\ur,\vr)$ is the resting state, $\ur=\vr=0$ for FHN kinetics and
$\ur=\ale/5$, $\vr=1+\ale^2/25$ for LE kinetics. 

Initial conditions were set as $\u(\x,0) = \ur+\us\Heav(\icw-\x)$,
$\v(\x,0)=\vr$, to initiate a wave starting from the left end of the
domain. Here $\Heav()$ is the Heaviside function, and the wave seed
length was typically chosen as $\icw=2$ or $\icw=4$.  The interval
length $\L$ was chosen sufficiently large, say for the system
\eqtwo{RXD,FHN} it was typically at least $\L=350$, to allow wave
propagation unaffected by boundaries, for some significant time.

To characterize shape of the waves emerging in simulations and its
evolution, we counted significant peaks (wavelets) in the solutions as
the number $\n$ of continuous intervals of $x$ where $\u-\ur>0.1$.  In
some regimes, this number varied with time, as the shape of enveloped
changed while propagating. We also measured the speed of individual
wavelets as the speed of the fore ends of these intervals at short
time intervals. To estimate the group velocity, we considered the fore
edge of the foremost significant peak over a longer time interval,
covering several oscillation periods.

To compare the oscillatory front of propagating waves to the
linearized theory, we took the $v$-component of the given solution in
the interval and selected the connected area in the $(\x,\t)$ plane
where $|v(x,t)|<0.1$ ahead of the main wave.  We numerically fitted
this grid function $v(x,t)$ to \eq{fit} using Gnuplot implementation
of Marquardt-Levenberg algorithm. The initial guess for parameters
$\C$, $\decr$, $\cg$, $\waven$, $\x$, $\freq$ was done ``by eye''.
The fitting was initially on a small interval in time, smaller than
the temporal period of the front oscillations, and then gradually
extended to a long time interval, so that the result of one fitting
was used as the initial guess for the next fitting.

\section{Results}
\seclabel{results}

\subsection{Overview of wave types}

\dblfigure{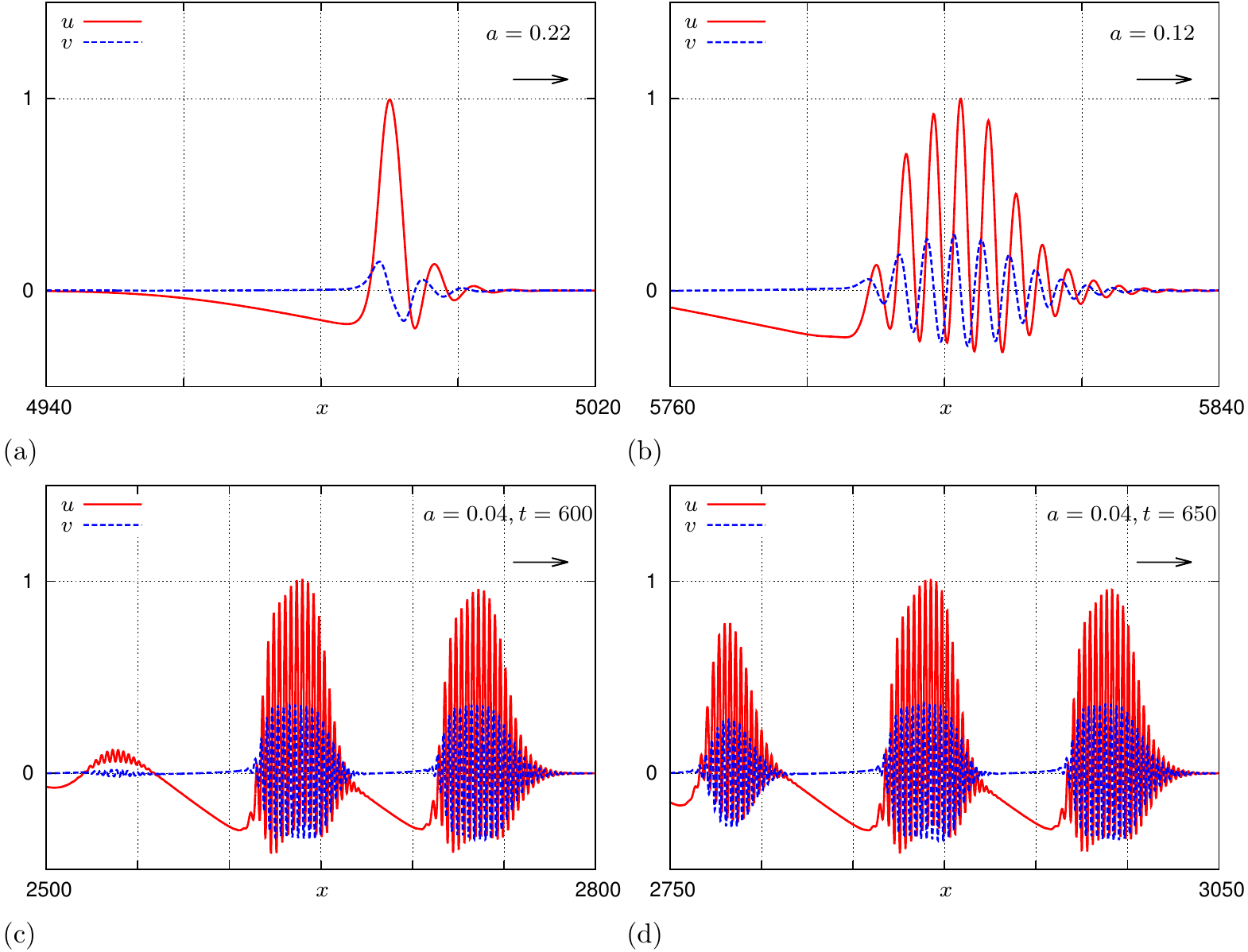}{%
  (color online) %
  Three typical wave regimes in the cross-diffusion
  system~\eqtwo{RXD,FHN} with $\ki=10$, $\eps=0.01$ for different
  values of $\a$. %
  (a) Simple quasi-soliton, $a=0.22$. %
  (b) Envelope quasi-soliton, $a=0.12$. %
  (c,d) Multi-envelope qausi-soliton, $a=0.04$, at two different time
  moments. %
}{threetypes}

\dblfigure{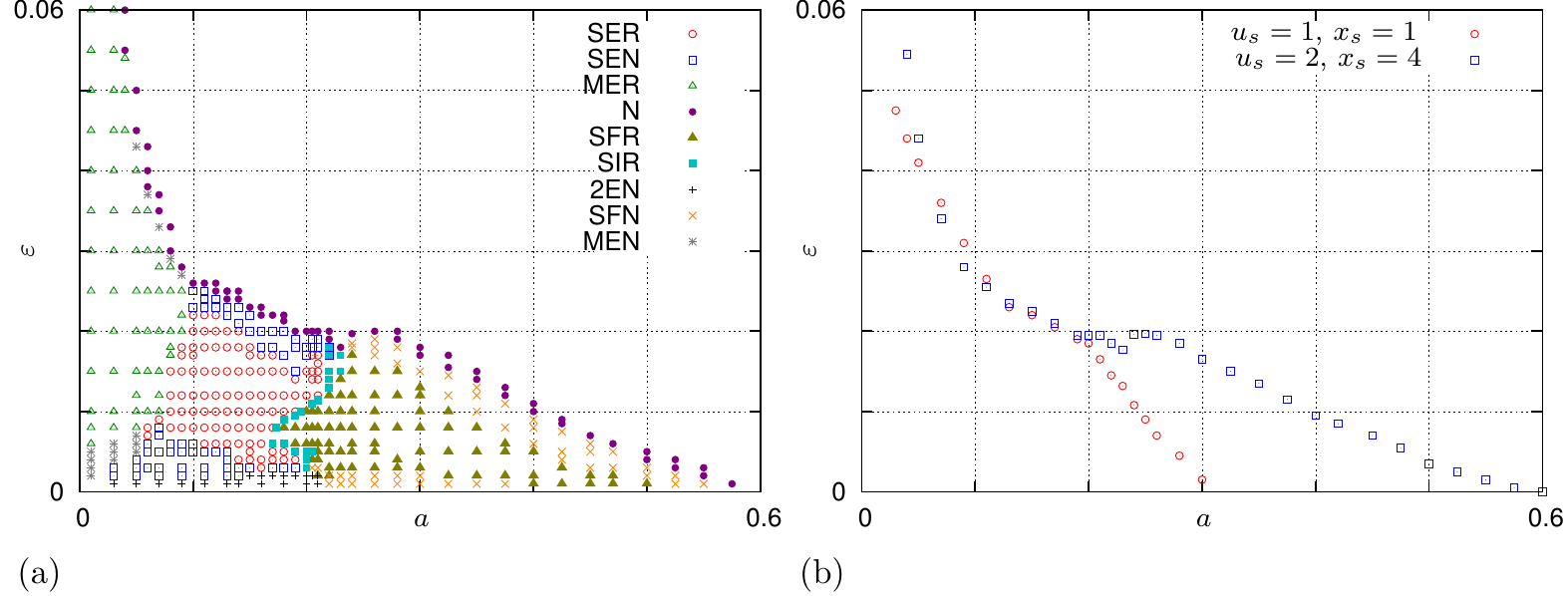}{%
  (a) The parametric regions corresponding to different wave regimes
  in \eqtwo{RXD,FHN} in the ($\a,\eps)$ plane at $\ki=10$, $\xs=4$,
  $\us=2$. %
  The abbreviations in the legend stand for various types of typical
  wave solutions: %
  SER single envelope reflecting; %
  SEN single envelope non-reflecting; %
  MER multiple envelope reflecting; %
  MEN multiple envelope non-reflecting; %
  SFR single fixed-shape reflecting; %
  SFN single fixed-shape non-reflecting; %
  SIR single intermediate (between single shape and envelope)
  reflecting; %
  2EN envelope non-reflecting with separate envelopes at the front and
  at the back with non-oscillating plateau between them; %
  N no propagation. %
  See Supplementary Material~\cite{epaps} for a movie. %
  (b) Boundaries of the regimes of propagation and decay (of any
  waves) for different initial conditions. %
}{fhn-parametric}

\Fig{threetypes} illustrates the three main types of waves in the
excitable cross-diffusion system~\eq{RXD} with FHN
kinetics~\eq{FHN}. \Fig{fhn-parametric} explains why these are
``main'' types. It shows the regions in the parametric plane
$(\a,\eps)$, and we see that the solutions shown in \fig{threetypes}
are represented by large parametric areas. Their common features are
quasi-soliton interaction and oscillatory front, and the differences
are in the propagation mode. A simple quasi-soliton
(\fig{threetypes}(a), abbreviation SFR in \fig{fhn-parametric}(a))
retains its shape as it propagates. A group, or envelope,
quasi-soliton (\fig{threetypes}(b), abbreviation SER in
\fig{fhn-parametric}(a)) does not have a fixed shape; instead it has
the form of spatiotemporal oscillations, whose envelope retains a
fixed unimodal shape as it propagates.  A multi-envelope quasi-soliton
(\fig{threetypes}(c,d), abbreviation MER in \fig{fhn-parametric}(a))
is shown at two time moments, to illustrate the dynamics of its
formation. At first, the emerging solution looks like an envelope
quasi-soliton; however after some time behind it forms another envelope
quasi-soliton, then behind that one yet another, and so it
continues. The interval of time between formation of new envelopes
depends on the parameters, e.g. it becomes smaller for smaller values
of $\a$.

\dblfigure{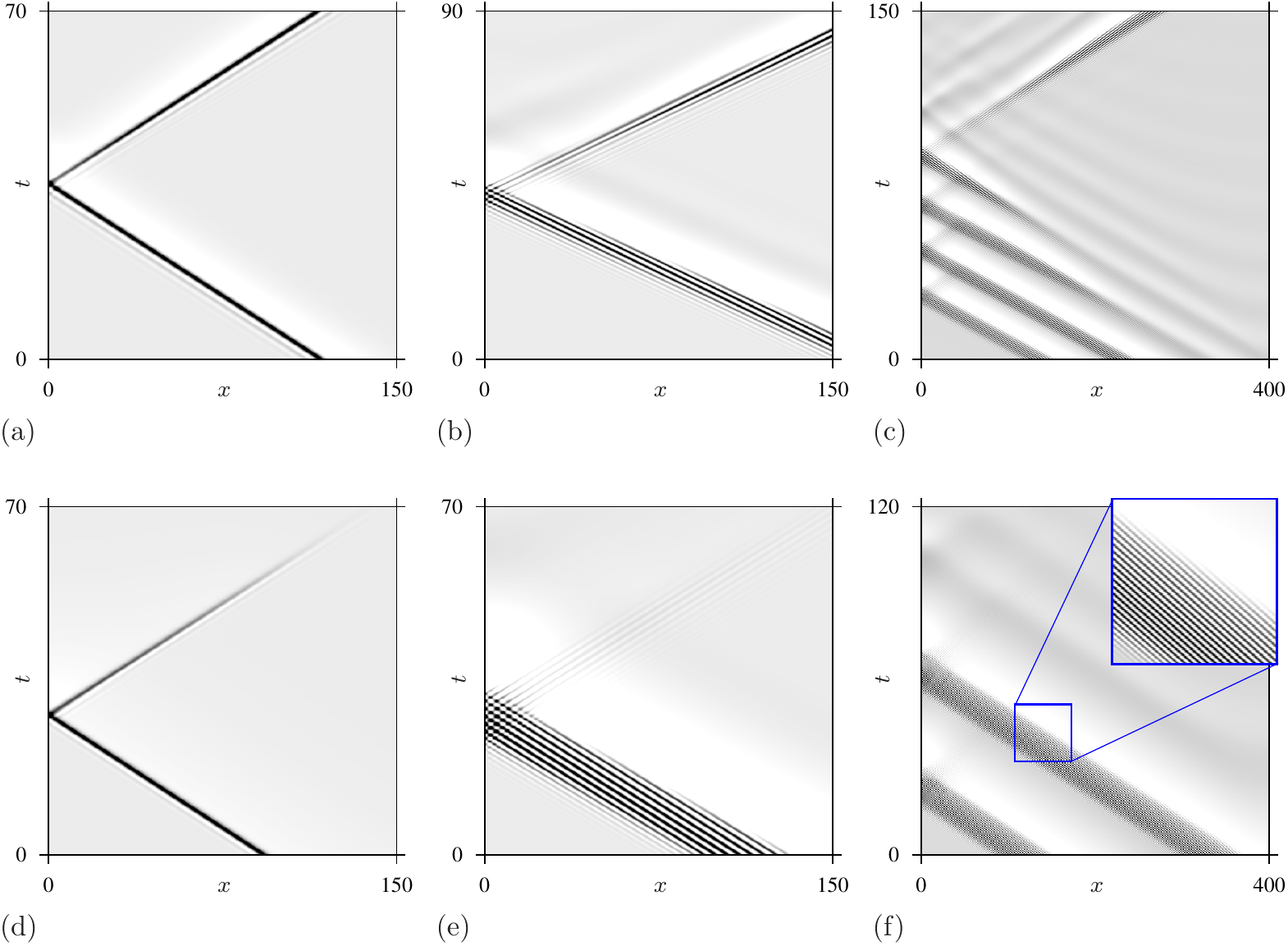}{
  Density plots of impact episodes for selected regimes designated in
  \fig{fhn-parametric}). %
  (a) SFR: single fixed-shaped (``simple'') quasi-soliton, $\a=0.22$, $\eps=0.01$. %
  (b) SER: single envelope quasi-soliton, $\a=0.1$, $\eps=0.01$. %
  (c) MER: multiple envelope quasi-soliton, $\a=0.02$, $\eps=0.01$. 
  In the panel, only the first reflected envelope has almost
  recovered within the view; other envelopes recover later. %
  (d) SFN: single fixed-shape non-reflecting wave, $\a=0.45$, $\eps=0.004$. %
  (e) SEN: single envelope non-reflecting wave, $\a=0.1$, $\eps=0.004$. %
  (f) MEN: multiple envelope non-reflecting wave, $\a=0.02$, $\eps=0.004$. %
  In (c) and (f), individual wavelets are not distinguishable at 
  printing resolution so only the envelope is in fact seen; in (f) the
  fine structure of the wavelets is shown magnified in the inset. %
  White corresponds to $\u=-0.3$, black corresponds to $\u=1$. Time
  reference point $t=0$ is chosen arbitrarily at the beginning of the
  selected episode; point $x=0$ corresponds to the left boundary of
  the interval. All simulations are done for $\dx=0.1$, $\dt=0.001$,
  $\L=400$, $\ki=10$. 
}{density}

Each of the three types of quasi-solitons shown in~\fig{threetypes}
has a counterpart type of solutions of similar propagation mode, but
without the quasi-soliton property, i.e. not reflecting upon collision
(abbreviations SFN, SEN, MEN in \fig{fhn-parametric}(a)).  Density
plots of interaction of the three main types of quasi-solitons and
their non-soliton counterparts are shown in~\fig{density}. Note that
the non-soliton regimes do not show immediate annihilation upon the
collision. Rather, the process looks like reflection with a decreased
amplitude, and subsequent decay, see~\fig{density}(d-f).

Apart from the non-reflecting counterparts to the three main types,
there are also ``non-propagating'' counterparts, all of which are
denoted by N in~\fig{fhn-parametric}(a). These regimes correspond to
waves that are in fact formed from the standard initial conditions,
but then decay after some time. Naturally, the success of initiation
of a propagating wave does in fact depend on the parameters of the
initial conditions:~\fig{fhn-parametric}(b) shows how the region of
single quasi-soliton differs for two different initial conditions. This
is of course expectable for excitable kinetics.

\dblfigure{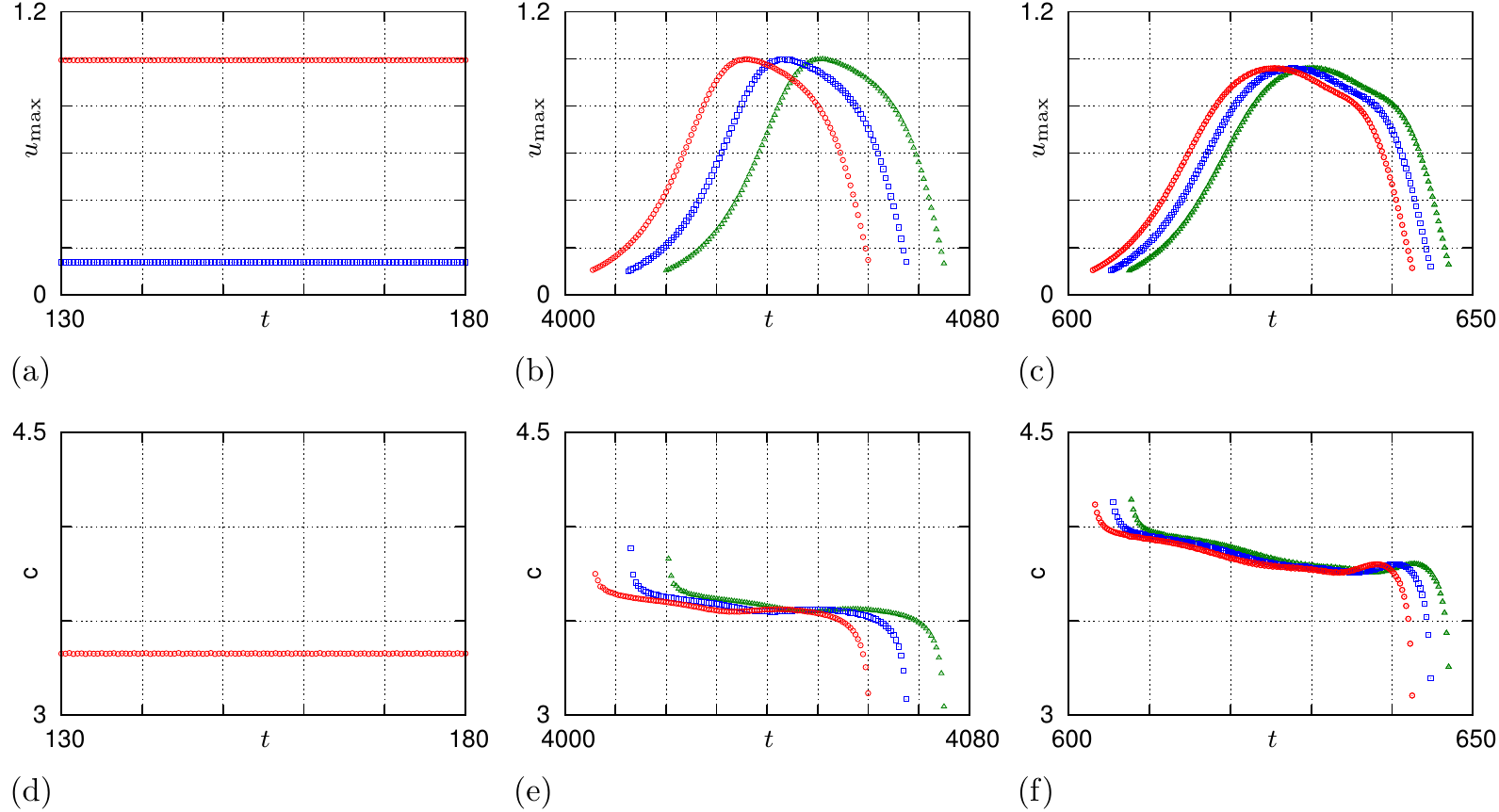}{
  Dynamics of (a--c) amplitudes and (d--f) velocities of individual
  wavelets for the three types of quasi-solitons in \eqtwo{RXD,FHN} for
  $\ki=10$, $\eps=0.01$. %
  (a,d) Simple quasi-soliton, $a=0.22$, see \fig{threetypes}(a). %
  (b,e) Envelope quasi-soliton, $\a=0.12$, see \fig{threetypes}(b). %
  (c,f) Multi-envelope quasi-solitons, $\a=0.04$, see
  \fig{threetypes}(c,d). %
}{ac-dynamics}

\dblfigure{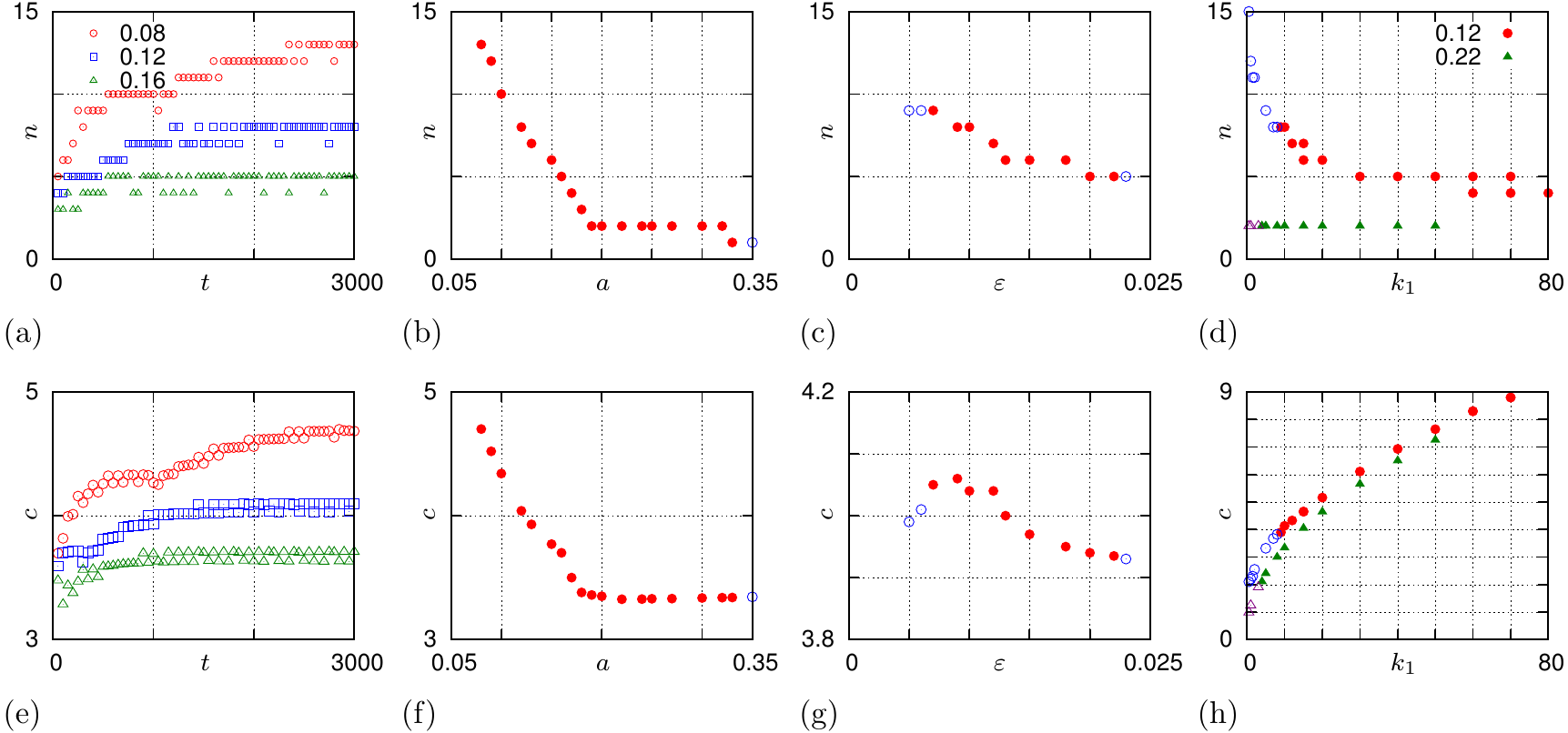}{%
  (a--d) Number of wavelets and %
  (e--h) velocities of the envelope waves in transient and in
  established regimes. %
  (a,e) The transients for $\ki=10$, $\eps=0.01$ and three values
  $\a=0.08$ (circles), $0.12$ (squares) and $0.16$ (triangles). %
  (b,f) Established quantities as functions of parameter $\a$ for
  fixed $\ki=10$, $\eps=0.01$. %
  (c,g) Established quantities as functions $\eps$ for fixed
  $\a=0.12$, $\ki=10$. %
  (d,h) Established quantities as functions $\ki$ for fixed
  $\eps=0.01$ and $\a=0.12$ (circles) and $\a=0.22$ (triangles). %
  In (b--d) and (f--h), filled symbols designate quasi-soliton (reflecting) waves
  and open symbols designate non-reflecting waves. %
}{nv-dynamics}

The analysis of the dynamics of the wavelets and wavespeeds for the
three main types of quasi-solitons, illustrated in \figs{ac-dynamics}
and \figref{nv-dynamics}, reveals:
\begin{itemize}
\item The amplitude and speed of the simple quasi-solitons do not
  change in time (\fig{ac-dynamics}(a,d)). 
\item For the envelope and multi-envelope quasi-solitons, the
  amplitudes of individual wavelets during their lifetime first grow
  to a certain maximum and then decrease monotinically
  (\fig{ac-dynamics}(b,c)). The speed of a wavelet (the phase velocity) is high at first,
  but the decreases non-monotonically (\fig{ac-dynamics}(e,f)).
\item In the process of establishment of an envelope quasi-soliton, the
  number of wavelets in it increases until saturation
  (\fig{nv-dynamics}(a)), and so does the speed of the envelope (the
  group velocity) (\fig{nv-dynamics}(e)). 
\item \Fig{nv-dynamics}(b,f) shows that in simple quasi-solitons ($a>0.2$), the
  number of wavelets remains the same ($\n=2$), and their
  speed remains approximately the same in that interval; whereas 
  in envelope quasi-solitons ($\a<0.2$), both the number of wavelets and their
  velocities increase with the decrease of $\a$. 
\item \Fig{nv-dynamics}(c,g) shows that increase of parameter $\eps$
  causes decrease of both the number of wavelets and of their speeds.
\item Parameter $\ki$ also plays a significant role in definining the
  wave regime and its parameters (\fig{nv-dynamics}(d,h)). 
\end{itemize}

\dblfigure{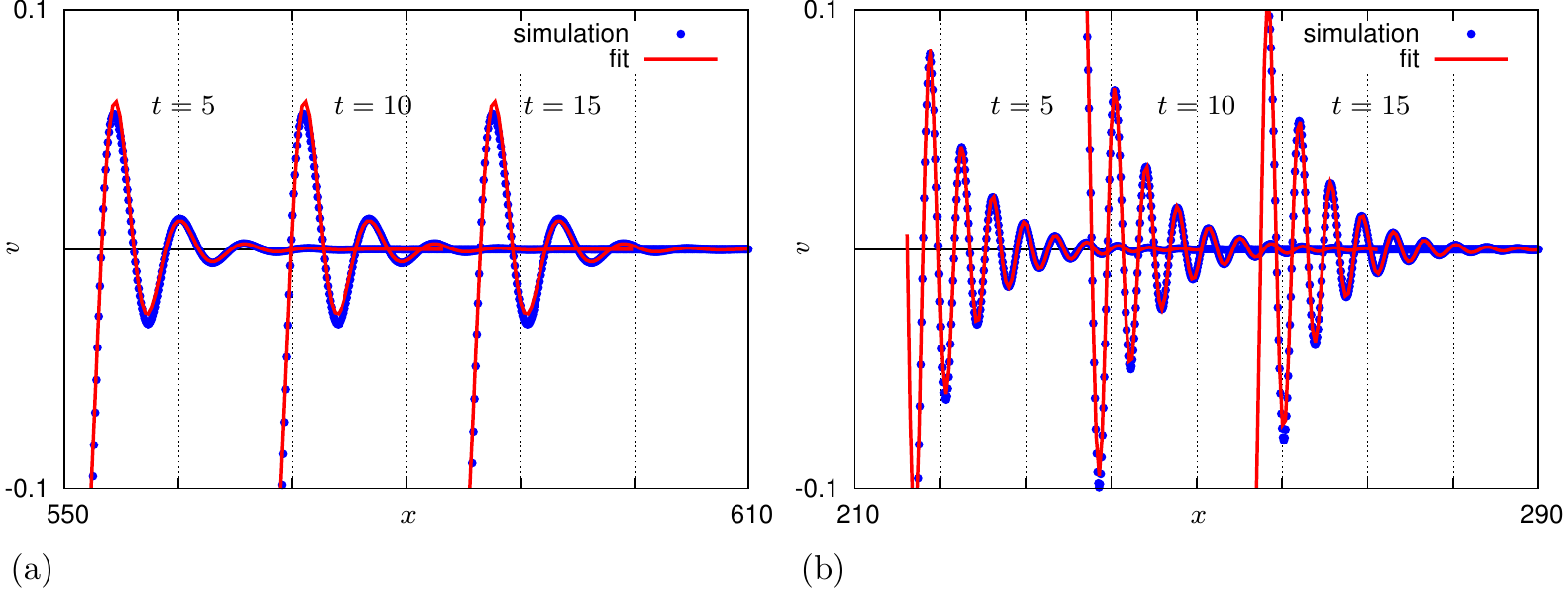}{
  Profiles of established propagating waves at selected moments of time for
  $\ki=10$, $\eps=0.01$, $\L=\infty$. The origins of the $t$ and $x$ axes
  are chosen arbitrarily. %
  (a) Simple quasi-soliton, $\a=0.22$. %
  (b) Envelope quasi-soliton, $\a=0.12$. %
}{profiles}

The oscillatory character of the fronts of cross-diffusion waves both
for simple quasi-solitons and for envelope quasi-solitons, which is
apparent from numerical simulations, is easily confirmed by
linearization of \eq{RXD} around the resting state.  The resting
states in both FHN~\eq{FHN} and LE~\eq{LE} kinetics are stable foci
which already shows propensity to oscillations. Taking the solution of
the linearized equation in the form
\begin{equation}
  \Mx{\u-\ur\\\v-\vr} \approx \Re{ 
    \C \evec \e^{-\decr(\x-\cg\t)} \e^{\i(\waven\x-\freq\t)} 
  },                                                        \eqlabel{fit}
\end{equation}
we need
\begin{equation}
  \A(\evalt,\evals)\evec=\mx{0}, \quad \evec\ne\mx{0}, \quad \det\A=0,
                                                            \eqlabel{disp} \\
\end{equation}
where
\begin{eqnarray*}
  &\A=\Mx{
    -\a-\evalt & -\ki+\evals^2 \\
    \eps-\evals^2 & -\evalt
  }, \\
  & \evalt = \decr\cg-\i\freq,\;\evals = -\decr+\i\waven.
\end{eqnarray*}
Equation~\eq{disp} imposes two constraints (for the real and imaginary
parts of the determinant) on the four real quantities
$\decr$, $\cg$, $\waven$ and $\freq$, so it is by far insufficient to
determine the selection of these parameters, but this equality can be
verified for the numerical simulations, in order to ensure that the
observed oscillatory fronts are not a numerical artefact but a true
property of the underlying partial differential equations. Hence we
fitted selected simulations around the fronts with the dependence
\eq{fit}. The quality of the fitting is illustrated by two examples
in \fig{profiles}. The fitted parameters satisfied \eq{disp} with good
accuracy; in both cases, they gave $|\det\A/(\Tr\A)^2|<10^{-3}$. 

Note that the approximation~\eq{fit} makes explicit the concepts of
wavelets (the oscillating factor $\e^{\i(\waven\x-\freq\t)}$), the
phase velocity (the ratio $\freq/\waven$), the envelope (in this case
the exponential shape $\e^{-\decr(\x-\cg\t)}$) and the group velocity
(the fitting parameter $\cg$). As expected, for the simple
quasi-soliton shown in~\fig{profiles}(a) the fitted group and phase
velocities coincided within the precision of fitting
($|\cg-\freq/\waven|<10^{-5}$). For the envelope quasi-soliton shown
in~\fig{profiles}(b) they were significantly different:
$\cg\approx4.077$, $\freq/\waven\approx3.586$.

\subsection{Multi-envelope quasi-solitons}

\dblfigure{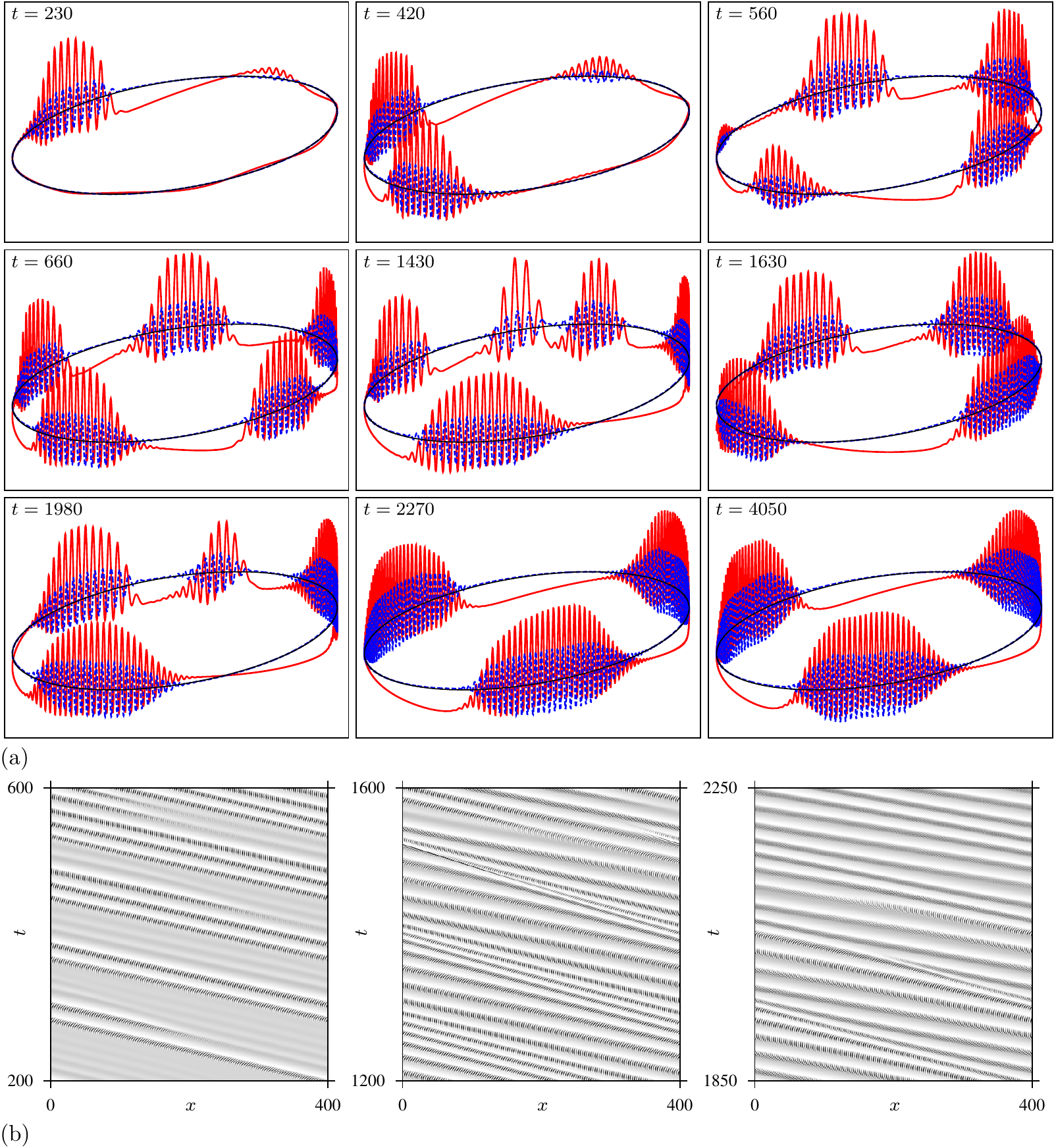}{
   Formation and evolution of a multi-envelope quasi-soliton regime on
   a circle (1D cable with periodic boundary conditions), for
   $\a=0.03$, $\ki=10$, $\eps=0.01$. (a) Snapshots of the profiles at
   selected moments of time. The waves and wavelets propagate
   counterclockwise. (b) Density plots of the $\u$ component of the solution for selected
   time intervals; white corresponds to $\umin=-0.2$, black
   corresponds to $\umax=1$
   corresponds to black.
   See also the movie in the Supplementary Materials~\cite{epaps}. 
}{multi}

We use the term multiplying envelope quasi-solitons (MEQS) to concisely
designate spontaneously multiplying envelope quasi-solitons. The
process of self-multiplication leads to eventually filling the whole
domain, behind the leading edge of the first group, with what appears
as a train of envelope quasi-solitons, i.e. a hierarchical,
quasi-periodic regime. This is illustrated in \fig{multi}(a) for
periodic boundary conditions, the setting that eliminates the
``leading edge'' complication mentioned above. One envelope
quasi-soliton (EQS) produced by the standard initial conditions
develops an instability at its tail, leading to generation of the
second EQS ($\t=230$).  The system of two EQSs generates a third
($\t=420$).  After forming of a system of five EQSs ($\t=600$), the
inverse transition happens, from five to four envelopes ($\t=1430$,
$\t=1630$), and then from four to three envelopes ($\t=1980$,
$\t=2270$), leading to an established, persistent state of three
envelopes ($\t=4050$). The same process is represented also as a
density plot in \fig{multi}(b).

\dblfigure{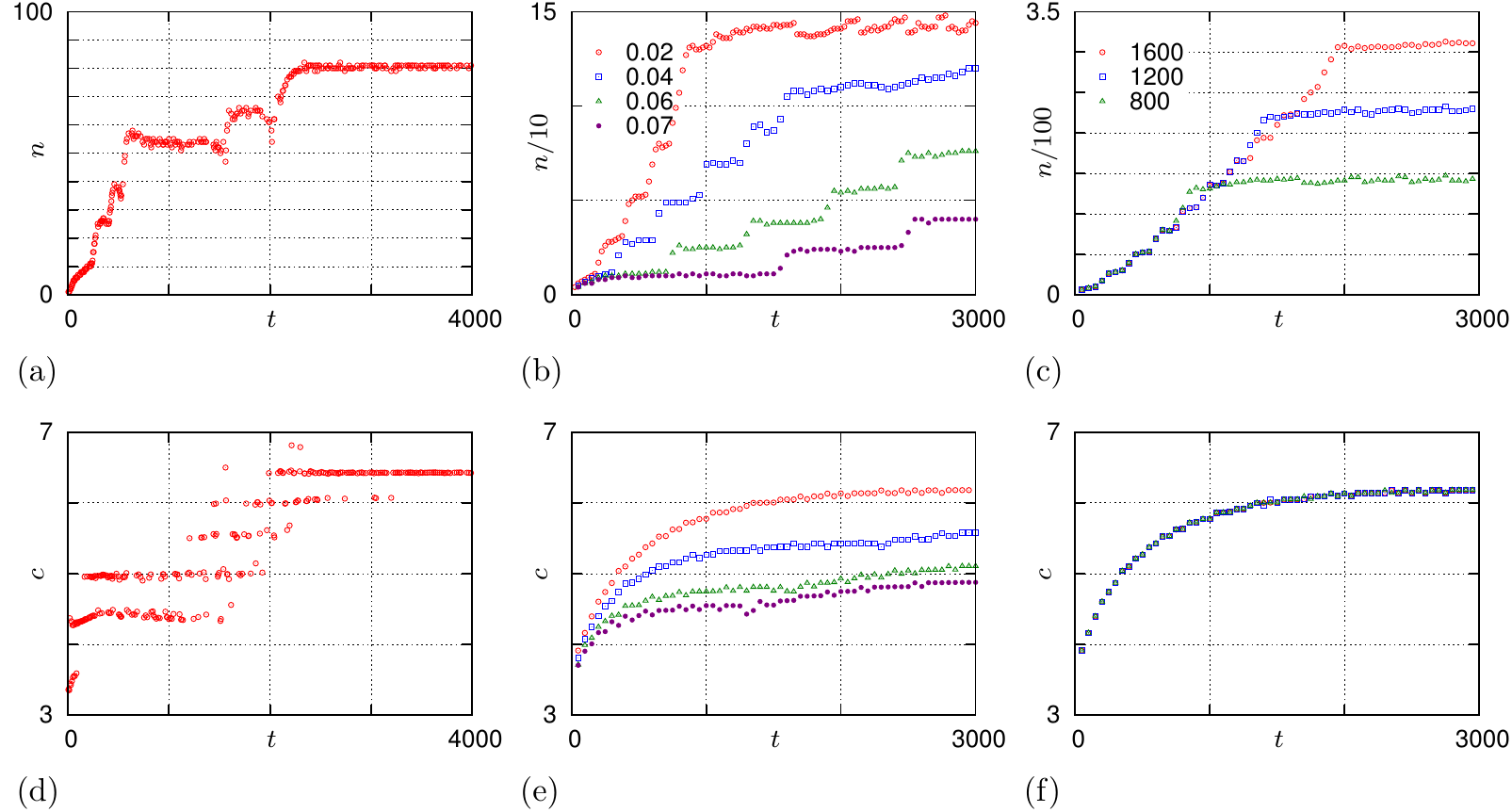}{%
  The dynamics of (a--c) numbers of wavelets and %
  (e--h) velocities of the envelope waves in multi-envelop
  quasi-soliton regimes. %
  (a,d) The quantities measured in a finite $\L$ with periodic
  boundary conditions; the same simulation as shown in \fig{multi}. %
  (b,e) The quantities are measured for the headmost $\L=800$
  interval of a wavetrain in an ``infinite length''
  computations, for $\eps=0.01$, $\ki=10$ and selected values of $\a$
  as indicated in the legend. %
  (c,f) The quantities are measured for $\eps=0.01$, $\ki=10$ and
  $\a=0.02$, with different measurement length $\L$ as indicated in
  the legend. In panel (f), the velocities measured for different $\L$
  are indistinguishable. %
}{multi-dynamics}

Panels (a,d) of \fig{multi-dynamics} analyse the dynamics of the
number of wavelets and the group (envelope) velocity for the
simulation shown in \fig{multi}. Both the wavelet number and the group
velocity grow, albeit non-monotonincally, till reaching stable
constant values, which corresponds to establishment of the stationary
regime of three envelopes shown in \fig{multi}. We stress that the
group velocity of the established multi-envelope soliton regime in a
circle is always higher than the speed of a similar regime on the
``infinite line'', which is illustrated in \fig{multi-dynamics}(b,e):
there the speed is established monotonically, and the number of
envelopes constantly increases. In \cite{tbi09} we have demonstrated
that in a cross-diffusion excitable system, the speed of a periodic
train of waves can be faster for smaller periods. There we called this
effect ``negative refractoriness'', meaning, using
electrophysiological terminology, that in the relative refractory
phase the excitability is enhanced rather than suppressed. In the
present case, we observe a similar negative refractoriness effect on
the higher level of the hierarchy, for envelope quasi-solitons (groups
of waves) rather than individual waves.

To conclude the analysis of the wavelet number and group speeds for
multi-envelope quasi-solitons, we note that for the MEQS on an
``infinite line'', as should be expected, does not depend on the
length of the interval used for computations, and the number of
envelope, obviously, does, see \fig{multi-dynamics}(c,f).

\dblfigure{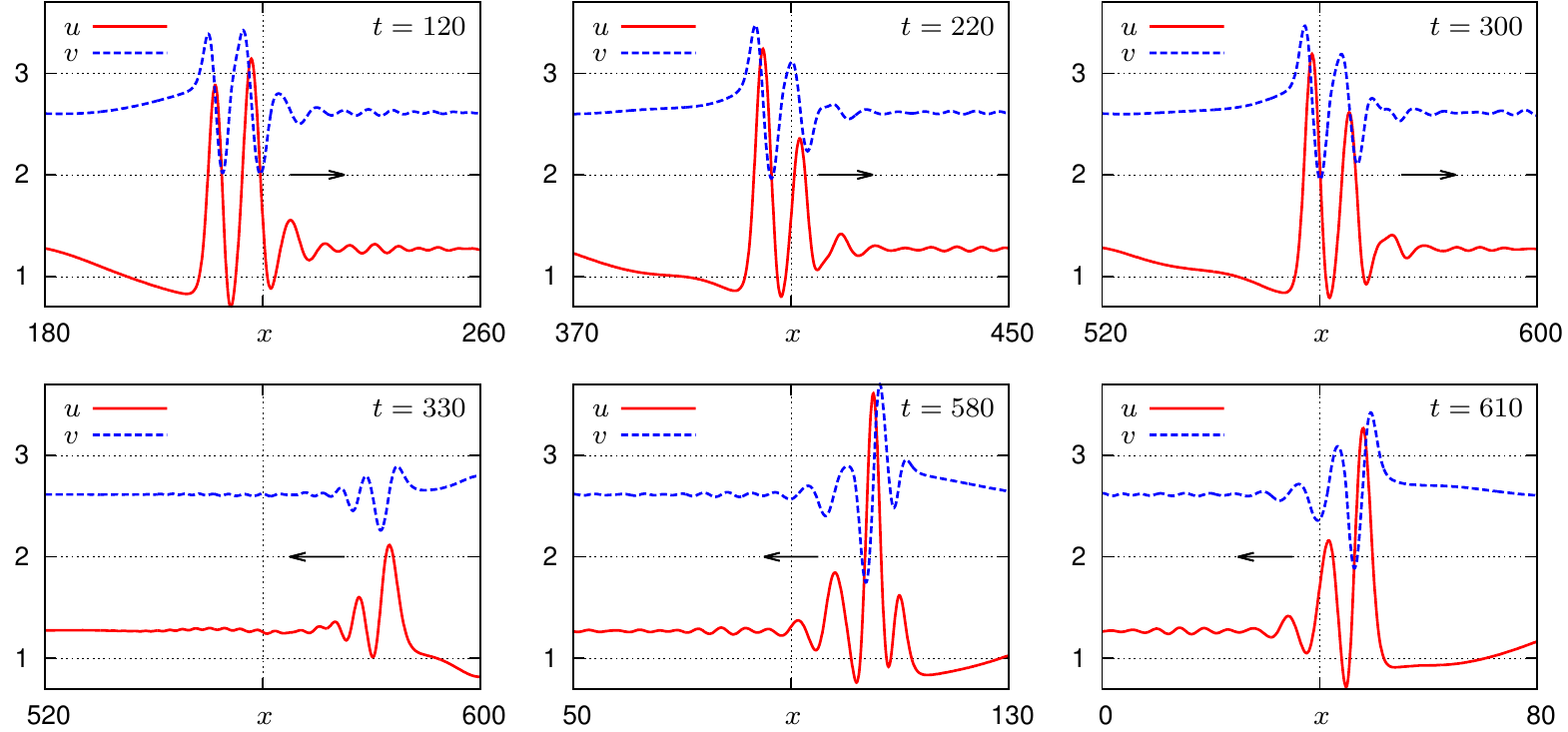}{%
  Quasi-soliton interaction of an envelope soliton with an impermeable
  boundary in LE model \eqtwo{RXD,LE},
  $\ale=6.35$, $\ble=0.045$. %
}{le-eqs}

\dblfigure{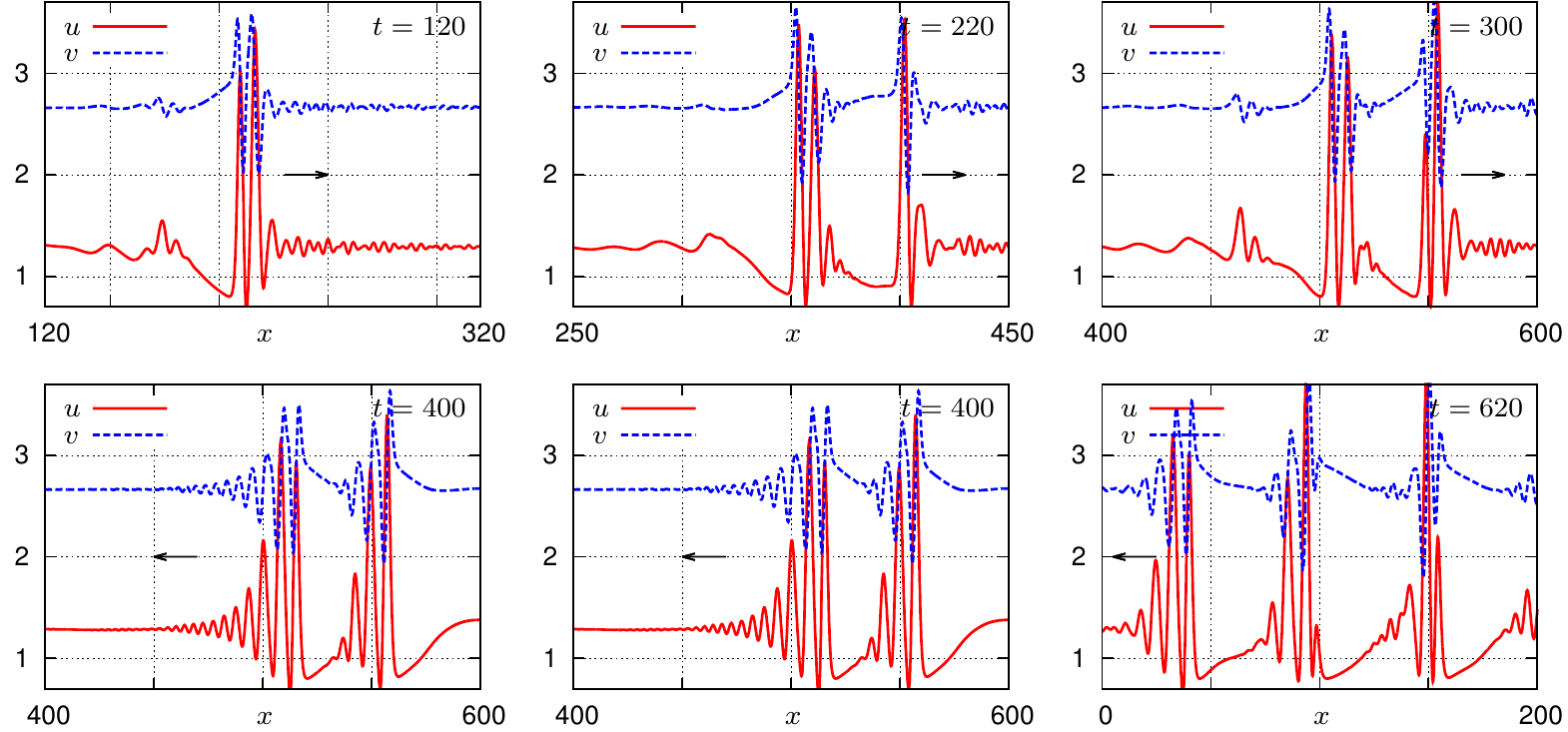}{%
  Formation of a multi-envelope quasi-soliton and its interaction with
  an impermeable boundary in LE model \eqtwo{RXD,LE},
  with $\ale=6.45$, $\ble=0.045$. %
}{le-meqs}

\sglfigure{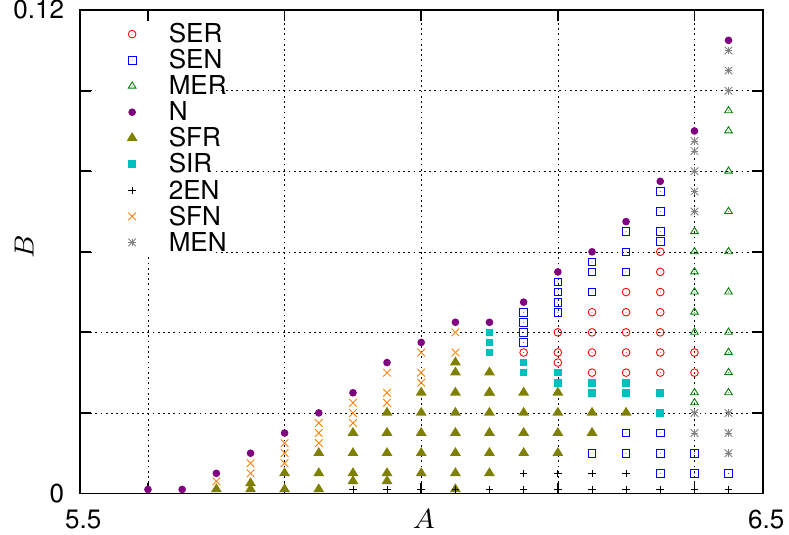}{%
  Parametric regions in the $(\ale,\ble)$ plane for different wave
  regimes in LE model \eqtwo{RXD,LE},.  The
  nomenclature of the wave regimes is the same as
  in~\fig{fhn-parametric}.%
}{le-parametric}

\subsection{Lengyell-Epstein kinetics}

Results of our numerical experiments with the
reaction--cross-diffusion system~\eq{RXD} with the LE kinetics~\eq{LE}
are qualitatively similar to those with the FHN kinetics~\eq{FHN},
described above. \Fig{le-eqs} illustrates the collision of an EQS with
an impremeable boundary for the LE kinetics. We can see that the
amplitudes of the wavelets decrease upon the collision ($\t=330$) and
then recover to their stationary values ($\t=580$,
$\t=610$). Similarly, \fig{le-meqs} illustrates formation of MEQS and
their interaction with the boundary for the LE kinetics. The
parametric portrait in the $(\ale,\ble)$ plane is shown in
\fig{le-parametric}. All the qualitatively distinct regimes identified
for the FHN kinetics and shown in \fig{fhn-parametric}, have been also
found for the LE kinetics and shown in \fig{le-parametric}.

\dblfigure{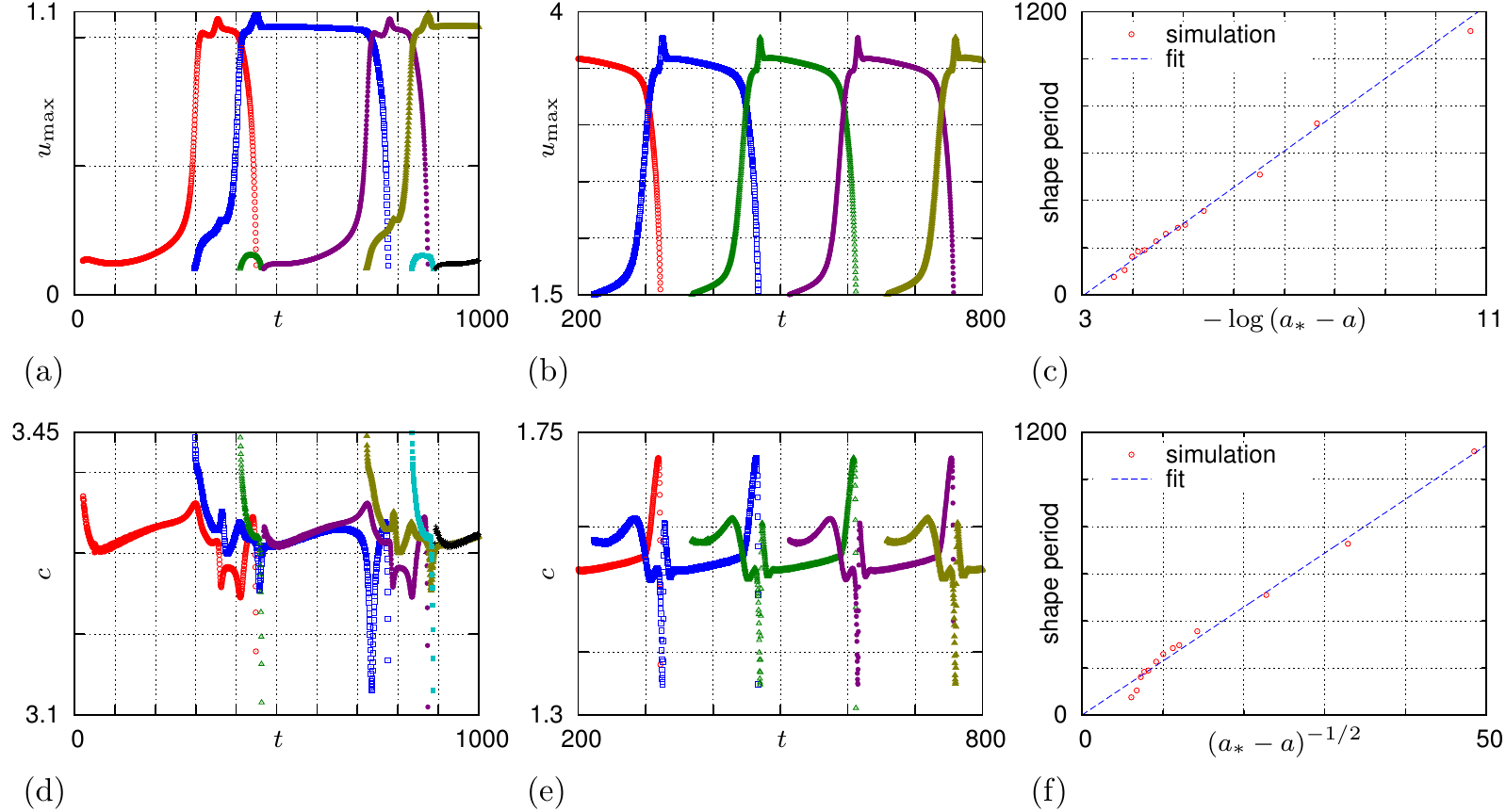}{
  Dynamics of quasi-solitons in the transitional area 'SIR' in
  \figs{fhn-parametric}(a) and \figref{le-parametric}. %
  (a,b) amplitudes and (d,e) velocities of individual wavelets of
  quasi-solitons; (a,d) FHN model \eqtwo{RXD,FHN}, $\a=0.18$,
  $\eps=0.006$. %
  (b,e) LE model \eqtwo{RXD,LE}, $\ale=6.25$, $\ble=0.025$. %
  (c,f) Dependence of the quasi-soliton shape change period in the FHN
  model as a function of parameter $\a$ at fixed $\eps=0.005$, and its
  best fits by the theoretical dependencies~\eq{ta-saddle} and
  \eq{ta-saddlefocus} respectively.  %
  See the last episode of Fig02 movie in the Supplementary
  material~\cite{epaps}. %
}{trans-dynamics}

\subsection{More exotic regimes}

Finally, we consider two more regimes to complete our overview.

The ``single intermediate reflecting'' (SIR) regime found both in
\fig{fhn-parametric}(a) and \fig{le-parametric} is ``intermediate'' in
the sense that it periodically changes its shape as it propagates, in
which sense it is similar to the envelope quasi-soliton; however most
of the time it propagates nearly as a simple quasi-soliton.  Only
during relatively short episodes, the wave undergoes transformation,
whereby it looses a wavelet at the tail and begets one at the front,
and these episodes are separated by relatively long periods when the
wave retains a constant shape. The dynamics of the parameters of such
a regime is shown in \fig{trans-dynamics}(a,b,d,e). This phenomenology
is reminiscent of a limit cycle born through bifurcation of a
homoclinic orbit. In our present context, this would of course be an
equivariant bifurcation with respect to the translations along the $x$
axis, or the bifurcation in the quotient system, i.e. the system
describing the evolution of the shape of the propagating wave, as
opposed to position of that wave (see~\cite{%
  Biktashev-etal-1996,%
  Biktashev-Holden-1998,%
  Chossat-2002,%
  Foulkes-Biktashev-2010%
}). Correspondingly, the limit %
cycle presents itself as the periodic repetition of the shapes of the
quasi-solitons, rather than periodic solutions in the usual sense.  In
the qualitative theory of ordinary differential equations, there are
two classical examples, which predict different dependencies of the
period on the bifurcation parameter. %
One is the bifurcation of a homoclinic loop of a saddle
  point~\cite{Shilnikov-1962-DM}; the other is the bifurcation of a
  homoclinic loop of a saddle-node~\cite{Shilnikov-1963-MS}, also
  known as SNIC (saddle-node in the invariant circle) bifurcation,
  SNIPER (Saddle-Node Infinite Period) bifurcation and ``infinite
  period'' bifurcation; see e.g.~\cite[Chapter 8.4]{Strogatz-2000}. In
  the case of a homoclinic of a saddle, the expected dependency is
\begin{equation}
  \Period \approx C_1 + C_2 \log\left(|\a-\ac|\right), \eqlabel{ta-saddle}
\end{equation}
where $\ac$ is the critical value of the bifurcation parameter $\a$
and $C_1$ and $C_2$ are some constants. For the bifurcation of
the homoclinic loop of a saddle-node, the asymptotic is different,
\begin{equation}
  \Period \approx C_3 |\a-\ac|^{-1/2}               \eqlabel{ta-saddlefocus}
\end{equation}
for some constant $C_3$.  The
fitting of the dependence of the soliton shape period on the
bifurcation parameter $\a$ in the FHN kinetics by \eq{ta-saddle} and
\eq{ta-saddlefocus} is shown in \fig{trans-dynamics}, panels (c) and
(f) respectively. In our case, the hypothetical limit cycles exist
for $\a<\ac$, and the best-fit bifurcation value for
\eq{ta-saddle} is $\ac\approx0.206477$, whereas for
\eq{ta-saddlefocus} it is $\ac\approx0.206925$. 

\dblfigure{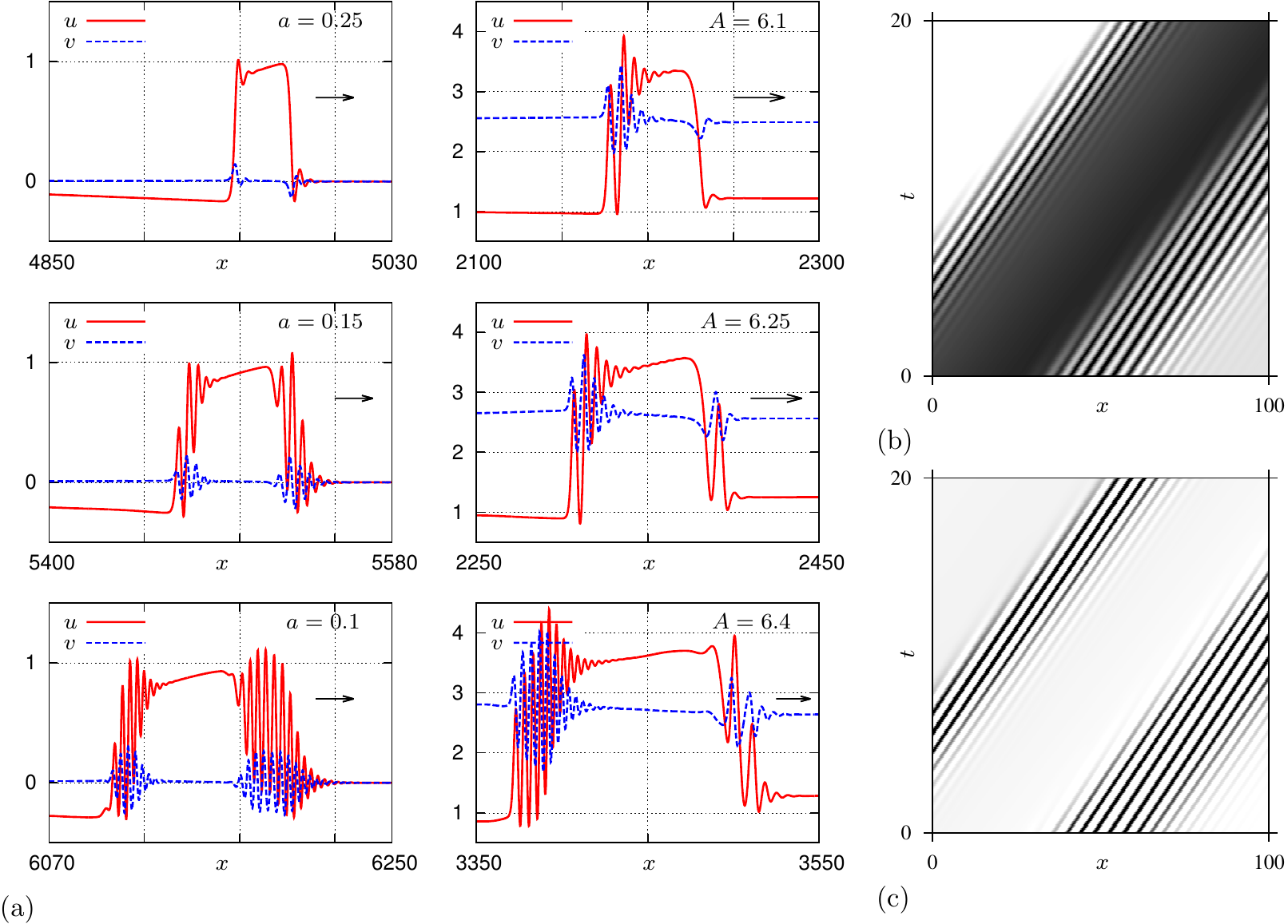}{
  (a) Profiles of established wave regimes for parameters from the `Z'
  zones in \figs{fhn-parametric}(a) and \figref{le-parametric}. %
  Left column: FHN model~\eqtwo{RXD,FHN} for $\eps=0.001$ and $\a=0.25$,
  $\a=0.15$ and $\a=0.1$ as shown by the legends. 
  Right column: LE model~\eqtwo{RXD,LE} for $\ble=0.001$
  and $\ale=6.1$, $\ale=6.25$ and $\ale=6.4$, as shown by the legends.  %
  (b,c) Fragment of a density plot of respectively the $\u$ and $v$ variables for the FHN
  model with $\a=0.1$. %
}{profdens}

\dblfigure{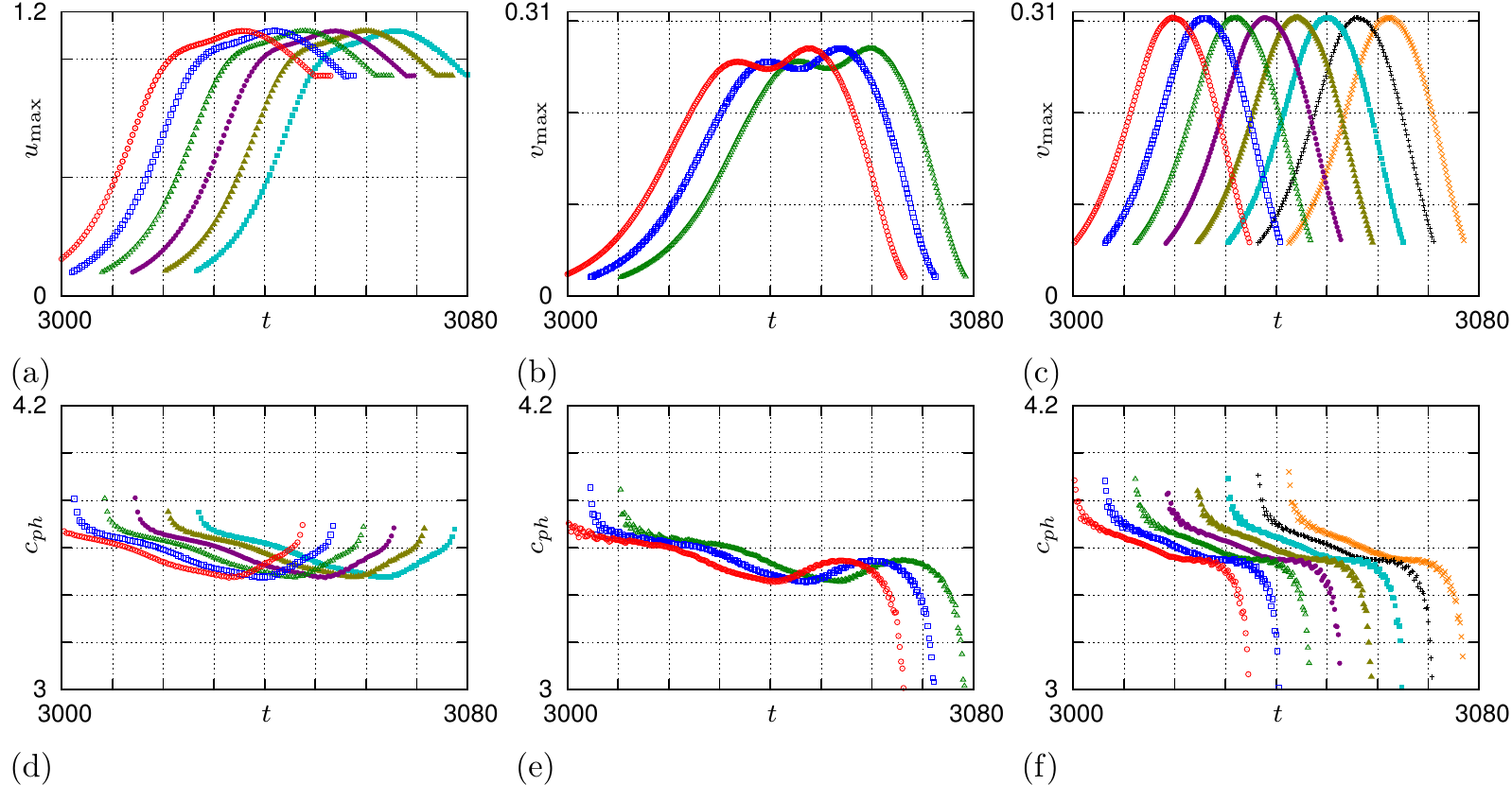}{
  Dynamics of the (a--c) amplitudes and (d--f) velocities of
  individual wavelets,
  (a,d) of the $\u$-variable in the front zone, (b,e) of the
  $\v$-variable in the front zone, (c,f) of the $\v$-variable in the
  back zone, for the FHN model~\eqtwo{RXD,FHN} at $\ki=10$, $\a=0.1$,
  $\eps=0.001$, see top bottom left panel of~\fig{profdens}(a). %
}{amp-dynamics}

The other regime is ``double-envelope non-reflecting'' (2EN) and it
has separate ``envelope'' trains at the front and at the back,
separated by a non-oscillating plateau, see \fig{profdens}.  The
corresponding dynamics of the wavelet amplitudes and their speeds is
shown in \fig{amp-dynamics}. This regime is observed for smaller
values of $\eps$ in the FHN kinetics (\fig{fhn-parametric}(a)) and smaller
values of $\ble$ in the LE kinetics (\fig{le-parametric}). 

\dblfigure{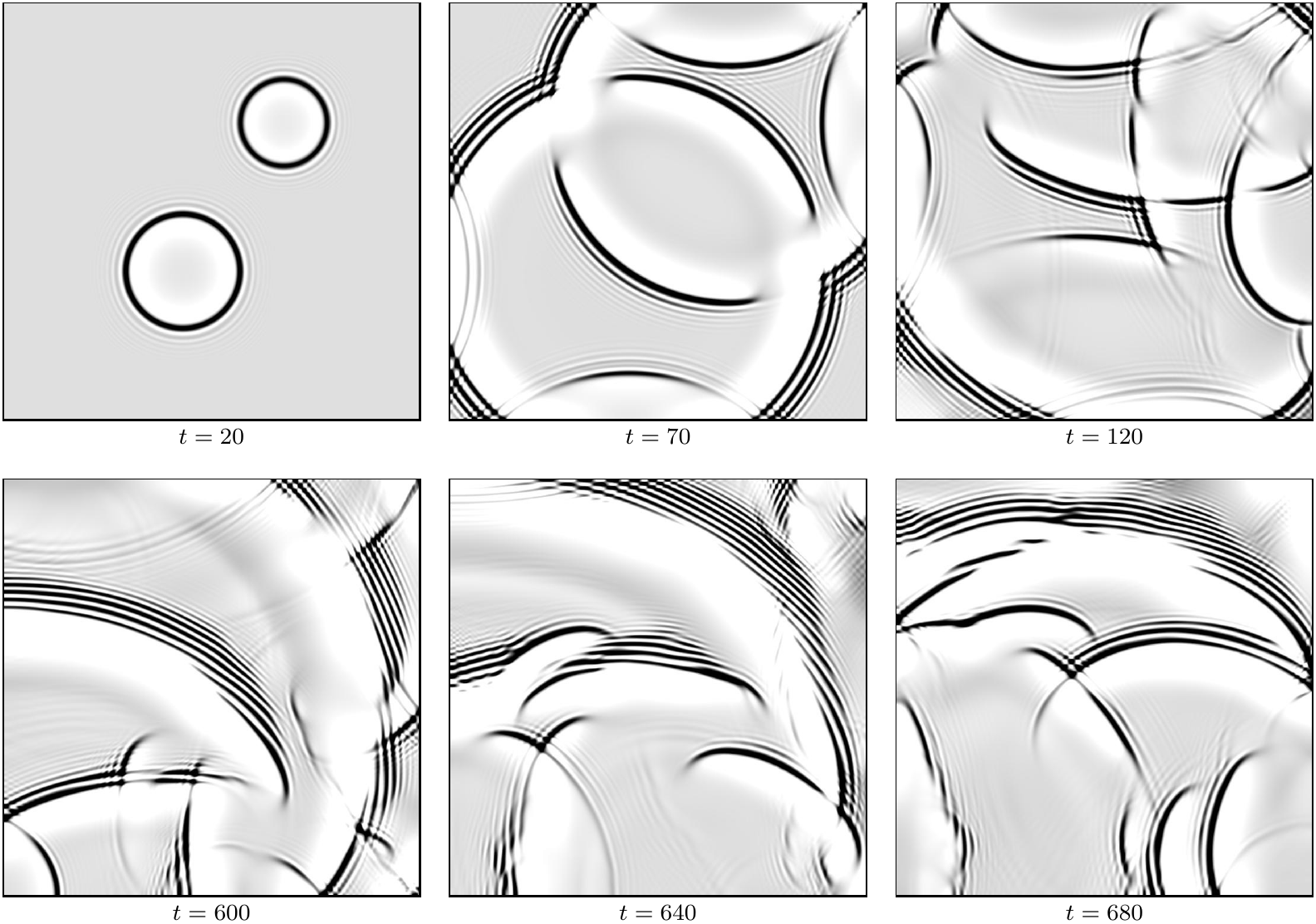}{
  Selected snapshots of ``wave flocks'' of envelope quasi-solitons in 2D FHN model with
  $\a=0.02$, $\eps=0.01$, $\ki=5$,
  $\hone=\htwo=0.3$, box
  size $140\times140$. %
  See Supplementary Material~\cite{epaps} for a movie. %
}{eqs2d}

\dblfigure{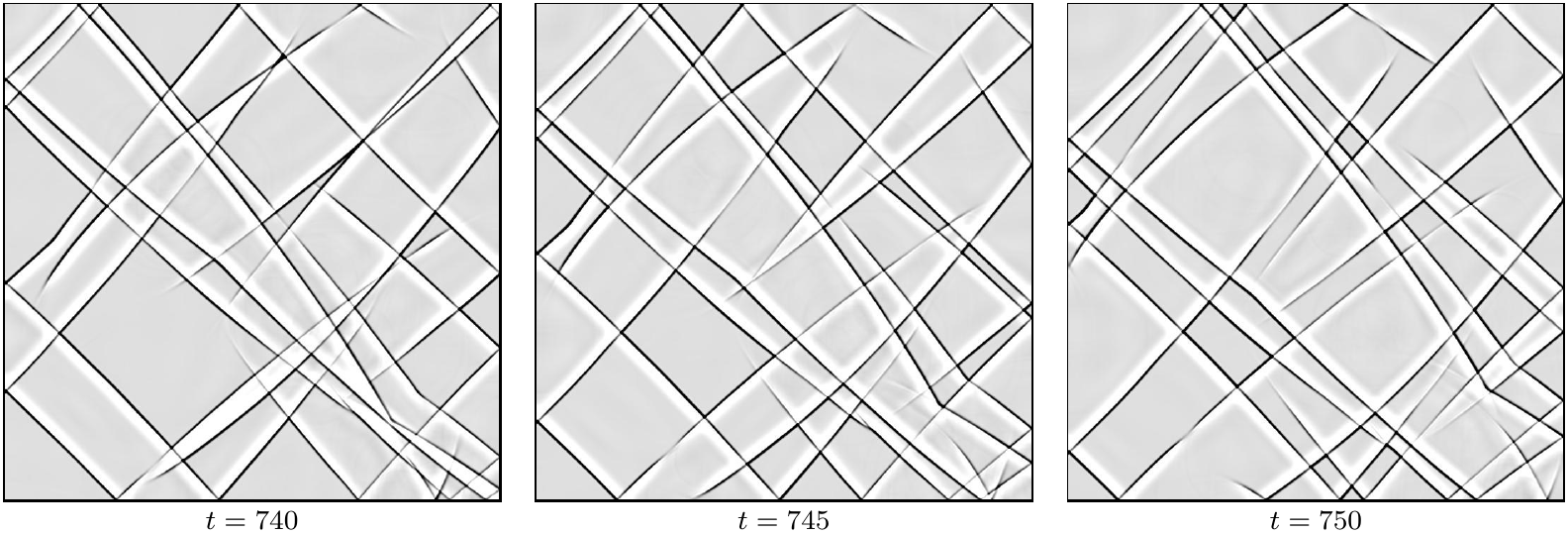}{
  Selected snapshots of ``wave grid'' of quasi-solitons in 2D FHN model with 
  $\a=0.02$, $\ki=30$, $\eps=0.01$, $\hone=\htwo=0.1$, box
  size $140\times140$. %
  See Supplementary Material~\cite{epaps} for a movie. %
}{qs2d}

\dblfigure{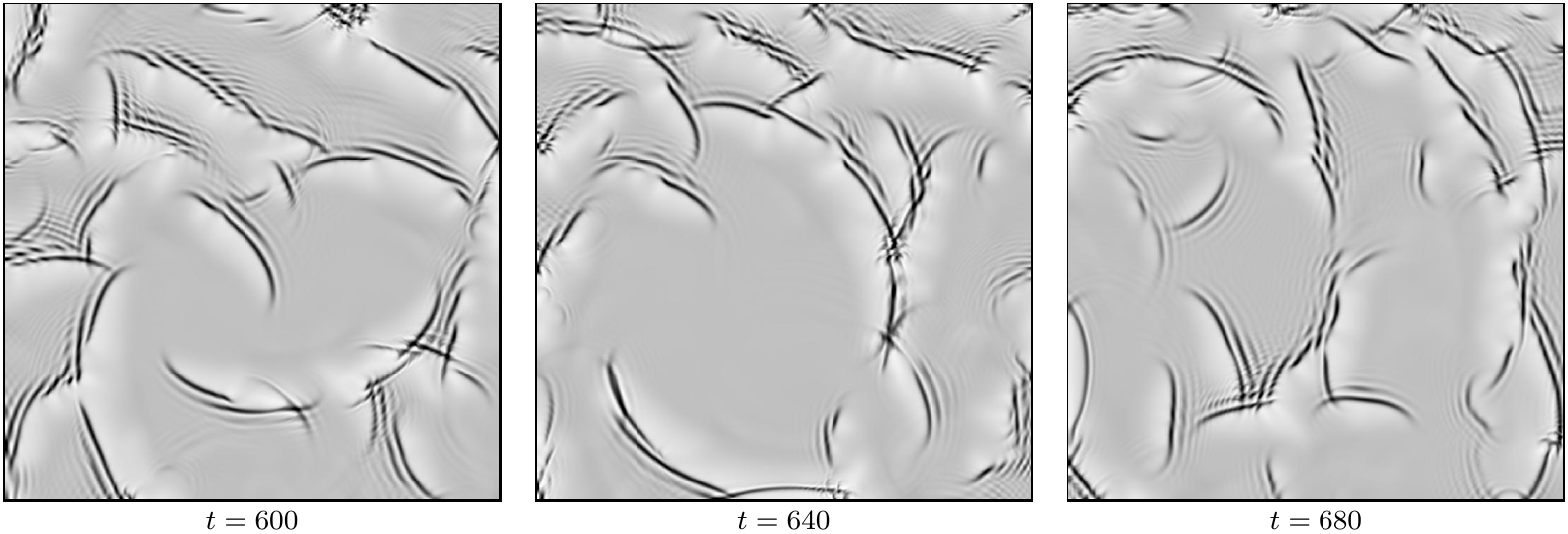}{
  Selected snapshots of ``wave flocks'' of envelope quasi-solitons in 2D LE model
  with $\ale=6.4$, $\ble=0.04$, $\hone=0.3$, $\htwo=0.3$, box size
  $140\times140$. %
  See Supplementary Material~\cite{epaps} for a movie. %
}{le-gang}

\subsection{Quasi-solitons in two spatial dimensions}

In \cite{QS3} we have shown that simple quasi-soliton waves in
two-dimensional excitable systems with cross-diffusion can penetrate
or break on collision. Whether the wavebreak occurs or not depended on
curvature and thickness of the waves, and also on the angle of their
collision, leading to emergence of complicated patterns.  The
two-dimensional extensions of the envelope and multi-envelope
quasi-solitons are no simpler; we present here only a few selected
examples, see \figs{eqs2d}--\figref{le-gang}.  The wavebreaks can
occur to whole wavetrains, as well as modify the number of a wavelets
in a train, and the result of a collision depends on the time interval
since a previous collision, so that encounters occuring in a quick
succession are more likely to lead to wavebreaks. This can lead to
``wave flocks'', that is, wave groups bounded not only lenthwise but
also sidewise, see~\fig{eqs2d} and \figref{le-gang}. For comparison,
\fig{qs2d} shows development of a ``wave grid'' of two-dimensional
simple quasi-solitons, i.e. the case where every wave has exactly one
wavelet; another reason for a different appearance is that the waves
at these parameters are more robust than those in \figs{eqs2d} and
\figref{le-gang}, and are broken less often, hence the typical sidewise
extent of the wave fragments is significantly longer.

\section{Discussion}
\seclabel{Discussion}

Solitons have attracted an enormous attention both from mathematical
viewpoint and from applications, ever since their discovery. For
applications, it has been always understood that the classical
solitons are an idealization, and it is therefore interesting to study
systems and solutions similar to solitons in different aspects and in
various degrees. Zakharov and
Kuznetsov~\cite{Zakharov-Kuznetsov-1998}, discussing optical solitons,
commented (translation is ours): \emph{%
  ``Objects called solitons in nonlinear optics are not solitons in the
  strict sense of the word. Those are quasi-solitons, approximate
  solutions of the Maxwell equations, depending on four parameters. Real
  stationary solitons, which propagate with constant speed and without
  changing their form, are exact solutions of the Maxwell equations,
  depending on two parameters.\ldots''%
}
We mention in passing that we are using the word ``quasi-solitons'' in
a different sense than~\cite{Zakharov-Kuznetsov-1998}; however, the main message is that
the completely integrable systems like nonlinear Schr\"odinger
equation are always an idealization and in real life one is interested
in broader class of equations and a broader class of solutions. 

The nonlinear dissipative waves in excitable and self-oscillatory
systems are traditionally considered an entirely different sort of
things from the integrable systems displaying the classical solitons:
the words ``active media'' and ``autowaves'' are sometimes also used
to characterize this different ``world''. 

The excitable media with cross-diffusion that we considered in this
paper are somewhat intermediate in that they present features in
common to both these different ``worlds''. On one hand, in a large
areas of parameters, we observe reflection from boundaries and
penetration through each other, although with a brief decrease, but
without change in shape and amplitude in the long run. The link to
dissipative waves is that in the established regimes have amplitude
and speed depending on the system parameters rather than initial
conditions.

In this paper, we have reviewed parametric regions and properties of a
few different regimes, such as simple quasi-solitons (corresponding to
classical solitons in integrable systems), envelope quasi-solitons
(corresponding to envelope, or group solitons, or breathers in
integrable systems). We have identified a transitional region between
simple and envelope quasi-solitons, which displays features of a
homoclinic bifurcation in the quotient system. We also have described
a regime we called multi-envelope quasi-solitons. This regime presents
a next level of hierarchy, after simple quasi-solitons (``solitary''
wave, stationary solution in a co-moving frame of reference) and
envelope quasi-solitons (``group'' wave, periodic solution in a
co-moving frame of reference), which are ``groups of groups of waves''
and apparently quasi-periodic solutions in a co-moving frame of
reference. One naturally wonders if this is the last level in thid
hierarchy or more complicated structures may be observed after a more
careful consideration --- however this is far beyond the framework of
the present study.

We have limited our consideration, with two simple exceptions, to a
purely empirical study, leaving a proper theoretical investigation for
the future. The two exceptions are that we confirm that the
oscillating fronts of the simple quasi-solitons and envelope
quasi-solitons observed in numerical simulations are in agreement with
the linearized theory, and that the periods of the quasi-solitons in
the transitional zone between simple and envelope are consistent with
a homoclinic bifurcation in a co-moving frame of reference. Further
theoretical progress may be achievable either by studying of the
quasi-soliton solutions as boundary-value problems by their numerical
continuation and bifurcation analysis, or by asymptotic methods.  At
present we can only speculate that one possibility is the limit of
many wavelets per envelope, which is inspired by observation that in
this limit the shape of the wavelets is nearly sinusoidal, so some
kind of averaging procedure may be appropriate in which the fast-time
``wavelet'' subsystem is linear and the nonlinearity only acting in
the averaged slow-time ``envelope'' subsystem. We have already
commented in~\cite{EQS} that treating cross-diffusion
FitzHugh-Nagumo system as a dissipative perturbation of the nonlinear
Schr\"odinger equation does not work out. A further observation is
that apparently this separation of time scales cannot be uniform, as some
parts of the envelope quasi-solitons that indeed look as
amplitude-modulated harmonic ``AC'' oscillations with a slow ``DC''
component, such as the head and the main body of the EQS illustrated
in~\fig{threetypes}(b), and the ``front'' and ``back'' oscillatory
pieces of the ``double-envelope'' regime shown in ~\fig{profdens}, and
some other parts which have only the slow component but no oscillating
component, such as the tail of the EQS of~\fig{threetypes}(b) and the
plateau and the tail of the double-envelope wave
of~\fig{profdens}. This suggests that any asymptotic description of
these waves will have to deal with matched asymptotics.

The systems we consider are not conservative, and the natural question
is where such systems can be found in nature. We have
  mentioned in the Introduction a number of applications that motivate
  consideration reaction-diffusion system with cross-diffusion
  components; a more extensive discussion of that can also be found in
  \cite{Vanag-Epstein-2009}.  Regimes resembling quasi-solitons and
  finite-length wavetrains phenomenologically have been observed in
  various places. The review~\cite{Vanag-Epstein-2009} describes a
  number of unusual wave regimes obtained in Belousov-Zhabotinsky (BZ)
  type reactions in microemulsions, including e.g. ``jumping waves''
  and ``packet waves'', which share some phenomenology with the group
  quasi-solitons. The ``packet waves'' are considered in more detail
  in~\cite{Vanag-Epstein-2002,Vanag-Epstein-2004}, which demonstrate,
  in particular, cases of quasi-solitonic behaviour of those,
  i.e. reflection from boundaries, see fig.~5(e)
  in~\cite{Vanag-Epstein-2004} paper---although it is difficult to be
  sure if it is the same as our group quasi-solitons as too little
  detail are given. Those packet waves have been reproduced in a model
  with self-diffusion only, but with three components. Another example
  of complicated wave patterns which may be related to quasi-solitons
  is given in \cite{Manz-etal-2006}, with experimental observations in
  a variant of BZ reaction as well as numerical simulation; again the
  simulations there were for a three-component reaction-diffusion
  system with self-diffusion only.
Regarding chemical systems, we
    must note that the models we considered here may not be expected
    to be realised literally. Apart from the choice of the kinetic
    functions and particularly of their parameters, based more on
    mathematical curiosity than real chemistry, the linear
    cross-diffusion terms as in~\eq{RXD} cannot describe real chemical
    systems as they do not guarantee positivity of solutions for
    positive initial conditions, so system~\eq{RXD} can only be
    considered as an idealization of~\eq{RT}, with corresponding
    restrictions. Further, our choice of the diffusivity matrix
    appears to be in contradiction with physical constraints related
    to the Second Law of Thermodynamics, which in particular require
    that the eigenvalues of the diffusivity matrix are real and
    positive, whereas ours are complex; see
    e.g.~\cite[p. 899]{Vanag-Epstein-2009} for a discussion. From this
    viewpoint, with respect to chemical systems, our solutions may be
    only considered as ``limit cases'', presenting regimes which
    possibly may be continued to parameter values that are physically
    realisable. On the other hand, it is well known that the
    aforementioned constraints apply to the \emph{actual} diffusion
    coefficients, whereas mathematical models obtained by asymptotic
    reduction deal with \emph{effective} diffusion coefficients, and
    the effective diffusion matrices may well have complex
    eigenvalues. A famous example is the complex Ginzburg-Landau
    equation (CGLE); see e.g.~\cite[Appendix B]{Kuramoto-1984}. This
    equation for one complex field, sometimes called ``order
    parameter'', emerges as a normal form of a supercritical Hopf
    bifurcation in the kinetic term of a generic reaction-diffusion
    system. This equation can also be written, in turn, as a
    two-component reaction-diffusion system, for the real and the
    imaginary parts of the order parameter. If the original
    reaction-diffusion system contains no cross-diffusion terms, but
    the self-diffusion terms are different, then the reduced
    reaction-diffusion system, corresponding to the CGLE, contains the
    full diffusion matrix including cross-diffusion term. Moreover, in
    that case the two eigenvalues of the diffusion matrix are
    complex. Incidentally, the two effective cross-diffusion
    coefficients will have signs opposite to each other, as in our
    Eq.~\eq{RXD}. 

  Speaking of other possible analogies found in literature, in
  nonlinear optics there is a class of phenomena called ``dissipative
  solitons'', which also could be related to our quasi-solitons. The
  literature on the topic is vast; we mention just one recent
  example~\cite{Korobko-etal-2014-OFT}; for instance, compare fig.~3
  in that paper with our~\fig{profdens}(a). Notice that the most
  popular class of models are variations of CGLE; e.g. models
  considered in~\cite{Korobko-etal-2014-OFT} involve effective
  diffusion matrices precisly of the form~\eq{RXD}.
  Propagating pulses of complicated shape, resembling group
  quasi-solitons, have also been observed in a model of blood
  clotting~\cite{Lobanova-Ataullakhanov-2004-PRL}. It is a
  three-component reaction-diffusion system, and ``muluti-hump''
  shapes are observed there for non-equal diffusion coefficients. A
  yet another possibility is the population dynamics with taxis of
  species or components onto each other, such as bacterial population
  waves; examples of nontrivial patterns there have been presented
  e.g. in~\cite{UFN07,Tsyganov-etal-1993}. A spatially extended
  population dynamics model considered in~\cite{Kuznetesov-etal-1994}
  does not present complicated waveforms but is interesting as it
  demonstrates emergence of cross-diffusion from a model with
  non-equal self-diffusion-only coefficients as a result of an
  asymptotic procedure. Finally we mention neural networks, where
  ``anti-phase wave patterns'', resembling group quasi-solitons, have
  been observed in networks of elements described by Morris-Lecar
  system~\cite{Dmitrichev-etal-2013}.  The question whether all these
  resemblances are superficial, or there is some deeper mathematical
  connection behind some of them, presents an interesting topic for
  further investigations. %

\section{Acknowledgements}

MAT was supported in part by RFBR grants No 13-01-00333 (Russia). VNB is
grateful to A.~Shilnikov and J.~Sieber for bibliographic advice and
inspiring discussions.

\end{document}